# A Cognitive Account of the Puzzle of Ideography

Commentary on "The Puzzle of Ideography" by Olivier Morin

Xerxes D. Arsiwalla[1,2],*

[1]Universitat Pompeu Fabra, Barcelona, Spain.

[2]Wolfram Research, Champaign, IL, USA.

*Email: x.d.arsiwalla@gmail.com

**Abstract**

In this commentary piece to "The Puzzle of Ideography" by Morin, we put forth a new cognitive account of the puzzle of ideography, that complements the standardization account of Morin. Efficient standardization of spoken language is phenomenologically attributed to a modality effect coupled with chunking of cognitive representations, further aided by multi-sensory integration and the serialized nature of attention. These cognitive mechanisms are crucial for explaining why languages dominate graphic codes for general-purpose human communication.

In Morin, 2022, the puzzle of ideography is broadly described as the near absence or rarity of self-sufficient generalist graphic codes for use in human communication. Morin proposes a resolution to this puzzle in terms of a "specialization hypothesis", which is subsequently explained in terms of a "standardization account". The latter does well in predicting empirical features in communication practice. Morin also pays due diligence to the alternative "learning account" explanation, which helps the discussion. However, both these explanations have primarily been cast in solely behavioral terms.

*Here, we will argue that there also exists a "cognitive account" which complements the behavioral explanation*. The cognitive account we propose is anchored on two well-studied cognitive phenomenon: (i) the *Modality Effect*, and (ii) *Chunking*. We show that (i) and (ii) offer a complimentary explanation to the specialization hypothesis.

Furthermore, combining the above two phenomena with two design features pertaining to cognitive processing: namely, *multi-sensory integration*, and the *serialized nature of attention*, explains what enables human languages to become self-standardizing, and also clarifies the precise role that learning plays.

The modality effect has extensively been studied in experimental psychology (Murdock, 1968; Penney, 1989). It refers to improved recall of items on a list when presented verbally in contrast to a purely visual representation. Subsequent studies, investigating this advantage of auditory over visual text modality during learning, have attributed the effect to factors including reduction in extraneous cognitive load, effective expansion of working memory and early sensory processes (Tabbers et al., 2001; Rummer et al., 2011). Additionally, the instructional benefits of presenting information across modalities found support in a meta-analysis involving 43 independent studies (Ginns, 2005).

Prima facie, the modality effect concerns learning and memory. However, one could argue that what can be learned or recalled better is also easier to repair during the process of turn-taking, and hence easier to standardize within a population. With respect to the specialization hypothesis, what this means is that at the population level, spoken forms of information exchange will dominate those that are purely visual. This comprises one part of our account.

*The other part relevant to the cognitive account is chunking.* In cognitive psychology, this refers to the process by which individual pieces of an information set are bound together into a meaningful whole (Miller, 1956; de Groot, 1968). Evidence for perceptual chunking is found in how primitive stimuli are grouped into larger conceptual groups, such as the manner by which letters are grouped into words, sentences or paragraphs (Johnson, 1970). Chunking is an effective strategy for overcoming capacity limits of working and long-term memory via coherent grouping of information (Laird et al., 1984). It has been observed in paradigms involving verbal learning, learning of perceptual-motor skills, expert memory, language acquisition and learning multiple representations (Gobet et al., 2001).

What is relevant here, is that chunking facilitates the creation of higher-order cognitive representations specific to the individual's perceptions and past experiences. Thus, allowing for greater abstraction and generality of cognitive representations. One could argue that this is not only about learning or memory, but underlies the core cognitive mechanism for generating flexible higher-order conceptualizations (see also Arsiwalla et al., 2023).

Why is this so important vis-a-vis the specialization hypothesis? The answer is that ideographs and other graphic codes that are not languages, allow for very limited abstraction and higher-order representation via compositionality of symbols. In language, compositions based on merely a limited set of letters, allows for creating an astoundingly large number of new concepts and higher-order representations. The same is not true for graphic codes (where base symbols already represent a specific concept). Take for example, the case of Bliss: composing symbols does not allow for reduction in complexity of expressions or creation of higher-order chunks. In contrast, a study presenting stimuli generated by an artificial grammar showed that subjects unintentionally learned to respond efficiently to the underlying structure (Servan-Schreiber et al., 1990). Moreover, it was demonstrated that the learning process was chunking and that grammatical knowledge was implicitly encoded in a hierarchical network of chunks.

*The point we want to emphasize is that the mechanism of chunking is precisely the brain's operationalization of compositionality of cognitive representations*. That the latter happen to be what languages are designed for, explains why the specialization hypothesis holds at all. Compositionality is the reason languages beat graphic codes for general-purpose communication. This also explains why writing is the only exception of a graphic code that is general-purpose: because it simply tags upon language.

Finally, we discuss two cognitive design features relevant to the puzzle. Firstly, multi-sensory integration. The effectiveness of a stimulus of one modality in eliciting attentive behaviors and faster reaction times was dramatically affected by the presence of a stimulus from another modality (Hershenson, 1962; Stein et al., 1989). The brain reinforces saliency of representations that use congruent evidence from multiple modalities. Thus, making the case for enhanced learnability of concepts involving unambiguous auditory and visual representation. This is true for words of a spoken language but not so for graphic codes which one may verbalizable in different ways and hence be difficult to standardize.

The second feature is the serialized nature of attention. This matters at the decoding / read-out stage. The brain's attention system seems designed to (consciously) process one task at a time (Baars, 1998). Studies have shown divided attention impedes performance (Pashler et al., 2001). Spoken language is strictly temporal and hence serialized. Written language tags along that temporality. Whereas, complex ideographs may have lay-outs as two-dimensional patterns, such that different decoders may assume a different read-out order. At the population level, this directly speaks to the difficulty in standardizing such codes.